\documentclass[12pt,reqno]{amsart}
\usepackage{graphicx,amsmath}
\usepackage[normalem]{ulem}
\usepackage{t1enc}              

\usepackage[utf8]{inputenc}    

\def\Mi{\mathbb{M}} 
\def\I{\mathbb{I}} 
\def\E{\mathbb{E}} 
\def\R{\mathbb{R}} 
\def\D{\mathbb{D}} 
\def\a{{\bar a}} \def\b{{\bar b}} 
\def\c{{\bar c}} \def\d{{\bar d}}  
\def\eh{{\bar e}} \def\fh{{\bar f}} 

\def\betag{Y} 	
\def\bg{y}			

\newcount\n  \newdimen\w

\def\Repeat#1#2{\n=#1\relax\loop\ifnum       
  \n>0\relax #2\advance\n by-1\repeat}

\long\def\OMIT#1{\relax }  

\def\re#1{(\ref{#1})}   
  
\def\eqn#1#2{ \begin{align} \label{#1}         #2 \end{align}}

\def\nl#1{          \\ \label{#1}        }  
\def\nnl#1{ \tag*{} \\ \label{#1}        }  

\def\delim#1#2#3{\csname\ifcase#1 relax\or   
   big\or Big\or bigg\or Bigg\fi\endcsname   
  {\ifcase#2\or\Delim#3\or\deliM#3\fi}}      
\def\Delim#1{\ifcase#1\relax\or(\or[\or\{\or<\or\langle\or|\or\|\or---{ }\fi}
\def\deliM#1{\ifcase#1\relax\or)\or]\or\}\or>\or\rangle\or|\or\|\or{ }---\fi}

\let\f\frac                     
        
\def\largerfrac#1#2#3{      
  \whichtypesize\n=\currenttypesize\advance\n by #1 \mathchoice
  {\setbox0\hbox{$\displaystyle-$} \w=.5\ht0\advance\w by-.5\dp0\setbox0
    \hbox{\typesize\n $\displaystyle-$} \advance\w by -.5\ht0\advance\w
    by .5\dp0\raise\w \hbox{\typesize\n$\displaystyle{\frac{#2}{#3}}$}}
  {\setbox0\hbox{$-$} \w=.5\ht0 \advance\w by -.5\dp0 \setbox0\hbox
    {\typesize\n $-$} \advance\w by-.5\ht0\advance\w by
    .5\dp0\raise\w\hbox{\typesize\n$\frac{#2}{#3}$}}
  {\setbox0\hbox{$\scriptstyle-$} \w=.5\ht0 \advance\w by-.5\dp0\setbox0
    \hbox{\typesize\n $\scriptstyle-$} \advance\w by -.5\ht0 \advance\w
    by .5\dp0 \raise\w\hbox{\typesize\n$\scriptstyle{\frac{#2}{#3}}$}}
  {\setbox0\hbox{$\scriptscriptstyle-$} \w=.5\ht0
    \advance\w by -.5\dp0 \setbox0\hbox{\typesize\n
    $\scriptscriptstyle-$} \advance\w by -.5\ht0 \advance\w by .5\dp0
    \raise\w\hbox{\typesize\n$\scriptscriptstyle{\frac{#2}{#3}}$}}  }

\begin{document}
	
	\title{Galilean relativistic fluid mechanics}
	\author{V\'an P.$^{1,2,3}$ }
	\address{$^1$HAS, Wigner Research Centre for Physics, Institute of Particla and Nuclear Physics, \\  
	and {$^2$Budapest University of Technology and Economics, Department of Energy Engineering},\\
	Montavid Thermodynamic Research Group}
	
	\date{\today}
	

\begin{abstract}
	Single component nonrelativistic dissipative fluids are treated independently of reference frames and flow-frames. First the basic fields and their balances, then the related thermodynamic relations and the entropy production are calculated and the linear constitutive relations are given.

	The usual basic fields of mass, momentum, energy and their current densities, the heat flux, pressure tensor and diffusion flux are the time- and spacelike components of the third order mass-momentum-energy density-flux four-tensor. The corresponding Galilean transformation rules of the physical quantities are derived. 

	It is proved, that the non-equilibrium thermodynamic frame theory, including the thermostatic Gibbs relation and extensivity condition and also the entropy production is independent of the reference frame and also the flow-frame of the fluid. The continuity-Fourier-Navier-Stokes equations are obtained almost in the traditional form  if the flow of the fluid is fixed to the temperature. This choice of the flow-frame is the thermo-flow.

	A simple consequence of the theory is that the relation between the total, kinetic and internal energies is a Galilean transformation rule.
\end{abstract}
	\maketitle
	
\section{Introduction}
	
	The concept of absolute, motion independent time is the result of our experience on the surface of the Earth, where the motions are slow and there is an apparently fixed background reference frame. However, the space is relative also in this case, it is different for different observers: the so-called non-relativistic space-time is Galilean relativistic. With the help of the notions of special relativistic space-time, the heuristic concepts of our everyday experience can be clarified and an exact mathematical model of Galilean relativistic space-time can be formulated \cite{Wey18bp,Hav64a,Fri83b,Mat86b,Mat93b,Ful08a2,MatVan07a}. 
	
	In spite of our seemingly self confident knowledge about slow motions of everyday life, such kind of exact space-time model is unavoidable to formulate and handle fundamental physical principles. If a physical phenomena happens without interacting with an observer, then one may find useful an adequate mathematical model, where this independent existence is reflected. Furthermore, considering the tempestuous history of concepts about space and time, this kind of model seems to be unavoidable.
	
	The mathematical notions in a physical theory are not simple tools, they are the building blocks, the material of the whole building. The shape and flexibility of blocks contributes to the stability of the construction. With properly formed blocks one can spare the mortar of imagination for a better use: for the design of the building and for other conceptual questions. A good example of the importance of proper building material may be elimination of space coordinates in classical field theories with the help of more abstract notions of vectors and tensors. Coordinate free concepts improve the focus on fundamental notions and are also helpful to develop better calculation methods in engineering.

	This work is an argument why reference frames should be avoided, and it provide tools how they can be eliminated,  similarly to coordinates, from theories of continua. Our example is single component fluid mechanics in this case. 

	Time is absolute in Galilean-relativistic space-time, it is the same for different observers, in other words, it is independent of reference frames. In the following the adjective {\em absolute} will be used in this sense for other reference frame free physical quantities, too. Other similar concepts, like objective or covariant, are mostly avoided.
	
	There are two points where our treatment is simplified. First, it  has been long known that space and time are not vector spaces but better represented by affine spaces, because none of them  have a canonical center \cite{Wey18bp,Hav64a,Fri83b}. In this work the affine character of Galilean space-time is not crucial, therefore we use	a vector representation for it. Second, we do not deal with the exact mathematical representation of physical units, nevertheless, it is interesting both from conceptual and practical points of view \cite{Dic62a, Bar96b,Lasz14m}. The definition of the complete space-time model is given in the Appendix. 
	
	{\em Absolute time is not a subspace of the Galilean relativistic space-time}. This is the critical, hidden problem of the usual superficial representations of Galilean relativistic space-time. At first it seems to be less important than the mentioned center and unit dependencies. However, if we represent space-time by a Cartesian product of the one dimensional time and three dimensional space, or even worse by $\R^4$, our theory is unavoidably reference frame dependent. The proper representation of the absolute time has consequences. 	One of them is, that in non-relativistic physical theories space-time four-covectors cannot be identified with four-vectors and they  transform differently when the reference frame changes. An other consequence is that the trace of a second order space-time four-tensor or a four-cotensor does not exist; only a mixed four-tensor has a trace. In this respect the Galilean relativistic space-time is not a specific low speed limit of special relativistic or general relativistic ones. The operations with space-time quantities are different from the usual relativistic operations, require care despite familiarity in special or general relativistic calculations. For example,  here some components of four-quantities, and not only the complete quantity, may be reference frame independent: the timelike component of a vector and the spacelike component of a covector is absolute.
	
	Several problematic aspects of non-relativistic physics are due to inappropriate space-time models:
	\begin{enumerate}
		\item {\em Principle of material frame indifference.} This principle formulates a simple and evident(looking) property, postulating  that the material must be independent of the reference frame, therefore the physical quantities, laws, equations of motions and material functions must reflect this requirement. In the usual formulations of the principle, the required invariance under various transformation rules is restricted to three dimensional spatial vectors and tensors. According to the usual explanations this restriction is due to the absolute time. For point masses one expects invariance for Galilean transformations related to inertial reference frames. For continua this is not enough, therefore invariance under rigid body motions of the reference frame is required. This most accepted formulation is due to Walter Noll (\cite{Nol58a,Nol67a,Nol04m,Nol06a,NolSeg10a}, see also in \cite{TruNol65b}). The principle and also its formulation initiated a long and unfinished discussion among those who are interested in the fundamental aspects of continuum physics. Here we give an incomplete list of the most important looking related works \cite{Zar903a,Jau11a,Old50a,Mul72a,EdeMcL73a,BamMor80a,Mur83a,Rys85a,Rys87c,Spe87c,Spe98a,SveBer99a,BerSve01a,Mas02a,Mur03a,Liu03a,Mur05a,Liu05a,YavEta06a,Fre09a,Mar05a,Mar07a,Mar08a,Mus98a,MusRes02a,MusRes08a,Mus12a}. 
		
		With the help of the Galilean relativistic space-time model one can show that the formal invariance (independency of the angular velocity of the rotating reference frame) is a wrong requirement, because reference frame independency is not always invariance, sometimes it requires that the characteristics of the relative motion appear in the transformation rules \cite{MatGru96a}, as it is evident in case of Galilean transformations. Moreover, from a space-time point of view the formulation of material frame indifference of Noll is self-contradictory \cite{MatVan06a}. 
		
		\item {\em Deformation measures.} Not only the constitutive functions, but also the basic fields of continuum mechanics are expected to be frame indifferent. One of the most discussed quantities in this respect is the finite elastic deformation. One can define  infinite number of deformation measures that are objective in the sense of Noll, and even more that are not. It is remarkable that the space-time requirements distinguish a single concept of deformation \cite{Ful08a3,FulVan12a,Ful15m}, which turned out to be advantageous from other points of view, too \cite{Eck48a,Bru01a,HorMur09a,NefEta14a,NefEta14m}.
		
		\item {\em Flow-frames.} A different problem concerns the basic field of fluid theory, the velocity of the fluid. According to a recent idea of Howard Brenner, the velocities in the mass, momentum and energy balances are not evidently the same. \cite{Bre05a,Bre05a1,Bre06a,Bre09a,Bre10a,Bre12a,Bre13a,Bre14a,BedEta06a,Ott07a}. This idea has a long history \cite{LanLif59b,DzyVol80a,Kli92a} and it is analogous to the problem of flow-frames in case of relativistic fluids \cite{Van09a,VanBir13p}. 
		What is the velocity of the fluid? What defines the motion from a physical point of view? The mass, the momentum, the energy or something else? Do we have a choice? In the usual fluid equations the velocities are relative velocities, therefore one must consider the role of the space-time to get a reliable answer.
			
		\item  {\em Compatibility with special relativity.} One may like to see the correspondence of physical quantities of Galilean relativistic fluids and special relativistic fluids beyond the usual comparison of relative quantities. In special relativity the basic quantity is the covariant second order energy-momentum density four-tensor. The energy is the time-timelike part of this tensor and the transformation properties of the energy are consequences of this fact. What could be a similar physical quantity in Galilean relativity? 
		
		\item  {\em The origin of transformation rules.} The energy, e.g. the kinetic energy, when expressed by the relative velocity is definitely relative and frame dependent. Is there an objective physical quantity behind the energy? Do we know the transformation rules of the non-relativistic energy?  What about the transformation rules of other physical quantities, like tensors and covectors?
		
		\item {\em Compatibility with kinetic theory.} The consistency with statistical physics, more properly with kinetic theory raises some questions, too. For example momentum series expansion of kinetic theory is informative regarding transformation rules of macroscopic quantities \cite{Rug89a,MulRug98b}. Moreover, the method of derivation of continuum equations (by Chapman-Enskog or momentum series expansion) is informative regarding the thermodynamics of fluids. For example the internal energy is determined in a close relation with the pressure. Is there an aspect in phenomenological continuum physics, where the hierarchical structure of momentum equations seems to be natural?  On the other hand, at the same time, nonrelativistic kinetic theory is considered as reference frame dependent, mostly due to macroscopic, phenomenological considerations (e.g. transformation properties of the heat flux \cite{MulRug98b}). This is a fundamental contradiction with general aspects of relativistic theories. 
		
		\item {\em Thermodynamics.} One may think, that thermodynamic relations are inherently absolute, because they express directly properties of materials. Covariance of thermodynamic relations is expected in the relativistic case, this is an important question since the birth of relativity theory (See the problem of temperature of moving bodies e.g. in  \cite{BirVan10a}). Interestingly, in the non-relativistic theory this question is rarely treated (but see e.g. \cite{KosLiu98a,Hor99a,Pri04a}). The Galilean-covariance of thermodynamical relations is not evident. Therefore it is not clear whether the Gibbs relation is independent of reference frames or not. One may expect also that dissipation is absolute, cannot depend on the motion of an external observer.
		
		\item {\em Natural philosophy.} Finally we emphasize that space-time formulation of a theory reorganizes our attitude to some fundamental concepts. In the framework of the Galilean relativistic space-time model reference frames, including inertial reference frames, are secondary, derived concepts. They are not fundamental, moreover should be avoided in case of general problems, contrary to the common belief \cite{Lan14a,Pfi14a}. 
	\end{enumerate}
	
	In the following we give a reference frame independent theory of single component dissipative Galilean-relativistic fluids starting with the absolute basic fields, their balances, the thermodynamic relations and finally calculating the entropy production. Along the absolute treatment the usual relative formulas and the corresponding transformation rules are calculated, together with the conditions that lead to the relative continuity--Navier-Stokes--Fourier system of equations.  
	
	In this work a particular version of abstract indexes of Penrose \cite{Pen07b} is introduced, that hopefully contributes to the transparency of the reference frame free meaning of the equations of Galilean relativistic fluid mechanics. We introduce three different indexes. The four-vectors and tensors of the Galilean relativistic space-time are denoted by upper $a,b,c,...$ indexes, the covectors and cotensors by lower $a,b,c,...$ indexes. Overlined indexes from the beginning of the alphabet, $\a,\b,\c,..$, denote spacelike four-vectors or spacelike parts of four-tensors, when positioned upper and spacelike four-covectors or four-cotensor components when positioned lower. It is important that the upper or lower position of the $a,b,c,...$ indexes is fixed, they cannot be pulled or pushed up or down, there is no canonical, observer independent identification between vectors and covectors. On the the hand the position of spacelike $\a,\b,\c,...$ indexes can be changed, because of the  Euclidean sturcture of space vectors. The indexes of the usual relative vectors and tensors, those that are related either  to the fluid or  the external reference frame are denoted distinctively by  $i,j,k,l,..$. These indexes are used whenever a single relative vector (typically the relative velocity) is present in the formula. The space-time model, the calculation rules and the notation are explained in the Appendices. The  following sections require the detailed knowledge of Appendix A, where the foundations of the Galilean-relativistic space-time model are given and also of Appendix B, where the most important calculation rules are derived and summarized.

	\section{Balances and their Galilean transformations}\label{f2}

	The fundamental balances of special relativistic fluids are expressed by four-divergences of the four-densities of the extensive physical quantities. In nonrelativistic physics these four-densities and four divergences are hidden, behind the usual relative formulation: the change of the extensive quantity is due to the local and simultaneous change inside the considered spatial region (timelike part) and the outward or inward flow (spacelike part). In the following we derive the usual relative forms of the balances in a Galilean relativistic framework.  Therefore, we express the four-vector field $A^a$ with the help of its relative parts due to the four-velocity field of the fluid $u^a$, introducing its $u$-form,  $A^a=Au^a+A^\a$, into the balance
	\eqn{A_bal}{
		\partial_aA^a =  D_u A + A\partial_a u^a +\partial_aA^\a = 
		D_uA+A\nabla_\a u^a + \nabla_\a A^\a=0,
	}
	where $a$ is a space-time index and $\a$ is the spacelike index. $A=\tau_aA^a$ and $A^\a= \pi^\a_{\;b}A^b$ are the timelike and $u$-spacelike parts of $A^a$, $D_u= u^a\partial_a$ and $\nabla_\a = \delta_\a^{\;b}\partial_b$ are the $u$-timelike and spacelike parts of space-time derivation  $\partial_a$ (see Appendix A).  The absolute balance, \re{A_bal} is expressed by the $u$-relative parts of the four-vector $A^a$ and four-covector $\partial_a$. 
	
	A relative velocity field plays a central role in the local and substantial relative balances of a fluid. This velocity is the relative velocity of the fluid and an inertial observer, therefore it is the difference of the four-velocity of the fluid, $u^a$, and the constant four-velocity field of an inertial external observer, $u'^a = const.$. The {\em local} form of the balance \re{A_bal} is obtained with the $u'$-form of the derivative and the four-vector field of the physical quantity:		
	\eqn{Al_bal}{
		\partial_aA^a =  D_{u'} A + A\partial_a u'^a +\partial_aA'^\a = 
		D_{u'} A +\nabla_\a A'^\a = 0.
	}
	
	The {\em substantial} form of the balance \re{A_bal} is obtained by substituting the four-velocity of the fluid, $u^a$, with the relative velocity of the fluid to the observer, $v^\a=u^a-u'^a$:
		\eqn{As_bal}{
		\partial_aA^a =  D_u A + A\partial_a u^a +\partial_aA^\a = 
		D_{u} A +A\nabla_\a v^\a + \nabla_\a A^\a = 0,
	}
	because  $u^a = u^a-u'^a +u'^a = v^\a + u'^a$ and $u'^a =\text{constant}$. In the local balances the time derivative is denoted as $D_{u'} = \partial_t$, in the substantial ones one may use a different notation, $D_u = d_t$, or the traditional dot, therefore $D_uA=d_tA=\dot A$. 
	
	In the following we apply a distinct notation, the indexes $i,j,k$, for spacelike relative vectors and covectors, in formulas related to an external observer. Both the  local  and the substantial balances \re{Al_bal} and \re{As_bal} can be written this way. The relative form of the  local balance is
	\eqn{Arl_bal}{
		\partial_t A +\nabla_i A'^i = 0,
	}
	and the substantial follows as
	\eqn{Ars_bal}{
		\dot  A +A\nabla_i v^i + \nabla_i A^i = 0.
	}
	It is easy to check that local balances can be obtained from substantial ones and vice versa with the Galilean transformation rules of covectors and vectors $\partial_t = \frac{d}{dt} -v^i\nabla_i$, and $A'^i = A^i+ A v^i$  (see appendix B).
	
	The substantial and local forms of the absolute balance depend on the components of the  four-derivative and the four-vector splitted by the four-velocities of the fluid, $u^a$, and the external observer, $u'^a$. It is important to remember, that the velocity field of the observer is given, the velocity field of the fluid is to be determined, we are looking for differential equations to determine it. The absolute equations are independent of {\em any} observer  \cite{MusRes08a}. 
	
	The fundamental physical information is given by the four-vector $A^a$, its four-divergence, and by the velocity field of the medium, $u^a$. Up to now, we did not say anything about the physical meaning of the four-velocity field of the medium. We will see, that the interpretation of the four-velocity field of the fluid requires thermodynamic considerations.
	
	However, first of all we should find the space-time representation of the physical quantities like mass, momentum, energy, heat flux, etc.. We will see, that actually we don not have several quantities, in Galilean relativity there is one, single, absolute physical quantity that characterizes a single component fluid.

	\subsection{Tensor or cotensor of how many orders?}

	In special relativistic fluid mechanics the physical quantities may be  scalars, four-vectors and also higher order objects, like the second order energy-momentum tensor. It is straightforward to assume that in Galilean relativistic space-time an analogous quantity exists like mass-momentum or energy-momentum.
	
	However, Galilean relativistic space-time is a more restricted framework than the special relativistic one, because the space-time vectors and covectors are related only by the linear structure, there is no Euclidean or pseudo-Euclidean structure on the space-time level which would admit their identification. Therefore, the divergence of a  four-covector or four-cotensor field does not exist independently of a reference frame. Similarly, there are no symmetric or antisymmetric parts of mixed four-tensors independently of reference frames. 
	
	On the other hand, the observed empirical, space-time related properties of relative physical quantities are reflected in their  transformation properties. In case of a reference and flow-frame independent Galilean relativistic theory these transformation properties can be deduced, they are consequences of the properties of space-time quantities. This theoretical framework must be harmonized with the empirical knowledge. For example, if the Galilean transformation rule of a physical quantity is like the transformation rule of the position, then this quantity maybe a spacelike part of a four vector. In this identification process the energy plays a distinguished role. One may come to the idea, that the relation of kinetic, internal and total energies  is a Galilean transformation rule. The total energy is the sum of internal and kinetic energies: from the point of view of a comoving reference frame, fixed to the flow of the medium, the energy is the internal energy, but for an external, inertial observer the kinetic energy have to be added. In a continuum the relation of the  total energy density $e_T$ and the density of the internal energy $e_b$ is the following:
	\eqn{etr0}{
		e_T= e_b + \f{\rho}{2}v^2.
	}
	due  two reference frames having  relative velocity  $v^i$. 
	
	These preliminary considerations outline the framework toward  understanding the relation of energy density and the space-time structure. In Appendix B we have calculated the transformation rules of four-vectors, four-covectors and different second order tensors. Scrutinizing the derived formulas one can observe, that a quadratic relative velocity in a transformation rule requires at least second order tensors. The time-timelike component of a second order four-tensor, the space-spacelike component of a second order four-cotensor and the time-spacelike or space-timelike component of a mixed second order tensor transforms quadratically with the relative velocity. However, the existence of balance form evolution equation restrict the possible choices, e.g. a second order four-cotensor field does not have a divergence in Galilean relativistic space-time, therefore cannot have a balance. However, energy may be also related to higher order tensors, like the time-timelike component of a third order, $\Mi\otimes\Mi^*\otimes\Mi^*$ valued mixed tensor field. 

	A further physical requirement is the compatibility to kinetic theory of gases. In the usual nonrelativistic theory the internal energy is the trace of the second order central momentum of the one particle probability distribution function by the relative velocity  \cite{MulRug98b,Lib90b}. Kinetic theory determines the energy in this respect, using also the ideal gas equation of state. From a space-time point of view an energy connected to the space-spacelike component of a second order four-tensor is not sufficient, because the energy balance requires energy flux, this way an additional tensorial order. In a Galilean relativistic continuum theory based on a second order four-tensor (the mass-momentum density tensor) one must introduce an independent, vectorial energy balance for the internal energy, therefore \re{etr0} cannot be obtained as a transformation rule \cite{Mat86a}. One may conclude, that in order to fulfill all these requirements, an $\Mi\otimes\Mi\otimes\Mi$ valued, third order tensor field emerges. The components of the four-divergence of this quantity must give the fundamental  balances of mass, momentum and energy together. In the next sections we will show that such a basic physical quantity leads to a consistent theory. 
	
	Astonishingly, one may obtain a rather similar theory assuming that the basic physical quantity is the above mentioned third order mixed tensor field. In both cases the transformation rules are the same and the entropy production is obtained in the same form (after a long calculation). The compatibility with the usual energy concept of kinetic theory compels us to consider in this paper a third order four-tensor as the basic physical quantity of a Galilean relativistic theory of single component fluids.
	
	Looking back to the relation \re{etr0} of total, internal and kinetic energies one may wonder about expected properties of transformation rules. Considering a third reference frame, it is easy to see, that the formula \re{etr0} is not a transitive rule. Our idea of an energy transformation rule is seemingly wrong. In the following we will see, that transitivity may be expected, our formula \re{etr0} is an oversimplified version of a more complicated relation. 

	\section{The mass-momentum-energy density-flux tensor and the related transformation rules}
	\label{ten}
	
	Let us consider the tensor field  $Z^{abc}: M \rightarrow \Mi\otimes\Mi\vee\Mi$, the {\em mass-momentum-energy density-flux third order four-tensor} of a single component fluid. We assume that the tensor field is symmetric in the second and third order, as it is indicated by the symbol $\vee$. In the following we do not explicitly denote this symmetry, $Z^{abc}=Z^{acb}$, only we  refer to it  if necessary. This tensor can be written in the following general $u$-form, with the components obtained by the four-velocity $u^a$:
		\eqn{TIE_3t}{
		Z^{abc} &= z^{bc} u^a + z^{\a bc}  \nnl{t1}
		& =\left(\rho u^b u^c + p^\b u^c + u^b p^\c + e^{\b\c} \right)u^a +
		\left(j^\a u^b u^c + P^{\a\b} u^c+ P^{\a\c}u^b + q^{\a\b\c}\right),
	}
	where
	\eqn{TIE_comp}{
		z^{bc} &= \tau_aZ^{abc}, \nl{tt1}
		z^{\a bc} &= \pi^\a_{\;d}Z^{dbc}. 
	}
	These components are the second order four-tensor of densities and the tensor of fluxes or current densities, that is $z^{bc}$ is the {\em mass-momentum-energy density tensor} and $z^{\a bc}$ is the {\em diffusion-pressure-heat flux} tensor. $\tau_a$ is the time evaluation and $\pi^\a_{\;b}$ is the $u$-projection for taking the $u$-spacelike parts of vectors. The further notations are:
	\begin{itemize}
		\item $\rho= \tau_b\tau_c z^{bc} =\tau_a\tau_b\tau_c Z^{abc}$ is the time-time-timelike part of the mass-momentum-energy density-flux tensor, the {\em mass density} or {\em density}. 
		\item $p^\b =\pi^\b_{\;d}\tau_c z^{dc}  = \tau_a\pi^\b_{\;d}\tau_c Z^{adc}$ is the time-time-spacelike part of the mass-momentum-energy density-flux tensor, the {\em momentum density}. Because of the symmetry of $Z^{abc}$ it is equal to the time-space-timelike part, $p^\c =  \tau_b\pi^\c_{\;d} z^{bd}$.
		\item $e^{\b\c} = \pi^\b_{\;d} \pi^\c_{\;e}  z^{de}  = \tau_a\pi^\b_{\;d} \pi^\c_{\;e} Z^{ade}$  is the {\em energy density tensor}, the time-space-spacelike part of $Z^{abc}$.
		\item $j^\a =\pi^\a_{\;d}\tau_b\tau_c Z^{dbc}$ is the {\em (self)diffusion flux}, the space-time-timelike part of $Z^{abc}$.
		\item $P^{\a\b} =\pi^\a_{\;d}\pi^\b_{\;e}\tau_c Z^{dec}$  is the {\em pressure}, the space-time-spacelike part of $Z^{abc}$. Because of the symmetry of the third order tensor it is equal to $P^{\a\c} =\pi^\a_{\;d}\tau_b\pi^\c_{\;e} Z^{dbe}$.
		\item $q^{\a\b\c} = \pi^\a_{\;d}\pi^\b_{\;e}\pi^\c_{\;f} Z^{def}$ is the {\em heat flux tensor}, the space-space-spacelike part of $Z^{abc}$.
	\end{itemize}
	
	The usual energy density and heat flux can be introduced according to the kinetic theory reducing the order of the corresponding tensors by two:
		\begin{itemize}
		\item $e = \f{1}{2} e^\a_{\;\a}$ is the {\em energy density},
		\item $q^{\a} = \f{1}{2} q^{\a\b}_{\;\;\;\b}$ is the {\em heat flux}.
	\end{itemize}
	
	All these quantities are defined  with the help of the velocity field of the medium, $u^a$. so they are independent of any external observers.

	\subsection{Transformation rules of time- and spacelike parts}
	
	The time- and spacelike components of a physical quantity obtained by an inertial observer of constant four-velocity $u'^a$ are relative quantities. The Galilean transformation rule of a physical quantity means expressing these relative quantities with the aid of the components corresponding to the four velocity $u^a$.
	In Appendix B the transformation rules of first and second order tensors are calculated by splitting the $u$-form of the tensors with the four-velocity $u'^a$ . The same procedure can be applied for the third order mass-momentum-energy density-flux tensor. In transformation formulas the relative velocity of the  fluid to the observer, $v^\a = u^a-u'^a$, appears naturally.
			
	The (mass)density, $\rho$, is a Galilean-scalar, it is invariant:
	\eqn{rk_tr}{
		\rho' = \tau_a\tau_b\tau_c Z^{abc} =  \rho.
	} 
	
	The momentum density and the mass density are components of an absolute four-vector, therefore the momentum density transforms like the spatial component of a four-vector:
		\eqn{pk_tr}{
		p'^ \b &= \pi'^\b_{\;d} \tau_c z^{dc} = 
		\pi'^\b_{\;\; d} \tau_c (\rho u^d u^c + p^\d u^c + u^d p^\c + e^{\d\c}) 
		=(\delta^\b_{\;\d} - u'^b\tau_d)(\rho u^d + p^\d) \nnl{pkt1}
		&= p^\b + \rho v^\b.
	}  
	
	The (self)diffusion flux transform like the momentum density, because together with mass density they are components of a four-vector, too:
	\eqn{jk_tr}{
		j'^\a &= j^\a + \rho v^\a.
	}
	
	The energy density is not a Galilean-scalar:
	\eqn{ek_tr}{
		e' &= \f{\delta_{\b\c}}{2}\pi'^\b_{\;\;d} \pi'^\c_{\;\;e} z^{de}  =     
		\f{\delta_{\b\c}}{2} (\delta^\b_{\;\d} - u'^b\tau_d)
		(\delta^\c_{\;\eh } -  u'^c\tau_e) 
		\left(\rho u^d u^e + p^\d u^e + u^d p^\eh + e^{\d\eh} \right) \nnl{ekt1} 
		&= \f{\delta_{\b\c}}{2} \left(\rho v^\b v^\c + p^\b v^\c + v^\b p^\c + 
		e^{\b\c} \right) = e+ p_\a v^\a + \f{\rho}{2} v_\a v^\a.
	}  
	
	The transformation rule of the pressure tensor is:
	\eqn{Pk_tr}{
		P'^{\a\b} &=\pi'^\a_{\;\;d}\pi'^\b_{\;\;e}\tau_c Z^{dec}= 
		(\delta^\a_{\;\d} - u'^a\tau_d)(\delta^\b_{\;\eh } -  u'^b\tau_e)
		\left(\rho u^d u^e + p^\d u^e + u^d j^\eh + P^{\d\eh} \right)
		\nnl{Pkt1}
		&= P^{\a\b} + p^\b v^\a + j^\a v^\b + \rho v^\a v^\b.
	}   
	The most complicated transformation rule belongs to the heat flux:
	\eqn{qk_tr}{
		q'^\a &= \f{\delta_{\b\c}}{2} q'^{\a\b\c} =  
		\f{\delta_{\b\c}}{2}\pi'^\a_{\;\;d}\pi'^\b_{\;\;e}\pi'^\c_{\;\;f} Z^{def} 	
		\nnl{qkt1}
		&=\f{\delta_{\b\c}}{2}  (\delta^\a_{\;\d} - u'^a\tau_d)
		(\delta^\b_{\;\eh} -  u'^b\tau_e)
		(\delta^\c_{\;\fh} - u'^c\tau_f)\left(
		\left(\rho u^e u^f + p^\eh u^f + u^e p^\fh + e^{\eh\fh} \right)u^d + 
		\right.\nnl{qkt2}
		&\qquad\qquad
		\left.\left(j^\d u^e u^f + P^{\d\eh} u^f + P^{\d\fh}u^e + 
		q^{\d\eh\fh}\right)\right)
		\nnl{qkt3}
		&= q^\a + (e + p_\b v^\b + \f{\rho}{2}v_\b v^\b)v^\a + 
		P^{\a\b}v_\b + j^\a \f{v^\b v_\b}{2}.
	}   
	
	Finally, we summarize these rules with the help of the usual 3-index notation:
	\eqn{rtr}{
		\rho' &=  \rho,  \nl{ptr}
		p'^i &= p^i + \rho v^i, \nl{jtr}
		j'^i &= j^i + \rho v^i, \nl{etr}
		e' &= e + p_iv^i + \f{\rho}{2}v^2, \nl{Ptr}
		P'^{ik} &=  P^{ik}+ \rho v^iv^k + p^kv^i + j^iv^k, \nl{qtr} 
		q'^i &= q^i + v^i\left(e + p_kv^k + \f{\rho}{2}v^2\right) + P^{ik}v_k +
		j^i \f{v^2}{2}, 
	}
	
	Two unusual quantities appeared in our reference frame free approach. One of them is the conductive current density of the mass, $j^\a$, the (self)diffusion flux. The other one is the momentum density, $p^\a$. One can reveal their role with the help of the absolute and relative balances of the fluid.

	\section{The fundamental balance of single component fluids and its components}
	
	The balances of mass, momentum and energy are the components derived from the divergence of the mass-momentum-energy density-flux tensor. The $u$-form of the divergence is:
	\eqn{Bask_bal}{
		\partial_a Z^{abc} &= \partial_a \left(z^{bc}u^a + z^{\a bc} \right) =
		{\dot z}^{bc} + z^{bc} \partial_a u^a +\partial_a z^{\a bc}  \nnl{bkb1}
		&= (\dot \rho u^c + \rho \dot u^c + \dot p^\c)u^b +
		(\rho \dot u^b+ \dot p^\b)u^c +
		p^\b\dot u^c+ p^\c\dot u^b + \dot e^{\b\c} +  \nnl{bb1}
		& \left(\rho u^b u^c+p^\b u^c + u^b p^\c +e^{\b\c} \right)\partial_au^a +
		u^b u^c\partial_a j^\a +j^\a u^c \partial_a u^b +j^\a u^b\partial_a u^c +
		\nnl{bkb2}
		& P^{\a\b} \partial_a u^c + u^c \partial_aP^{\a\b}+
		P^{\a\c} \partial_a u^b + u^b \partial_aP^{\a\c} +\partial_aq^{\a\b\c} = 
		0^{bc}.	
	}
	
	Here the dot denotes the $u$-time derivative: $u^a\partial_a( \ ) = D_u( \ ) = (\dot)$. The timelike part of the mass-momentum-energy balance \re{bkb2} is the mass-momentum balance:
	\eqn{mp_kbal}{
		\tau_c \partial_aZ^{abc} = \dot \rho u^b + \rho \dot u^b + \dot p^\b
		+ (\rho u^b + p^\b)\partial_au^a + u^a\partial_a j^\a + j^\a \partial_au^b
		+ \partial_a P^{\a\b} = 0^{b}.
	}
	
The timelike part of the mass-momentum balance (the time-timelike part of the mass-momentum-energy balance) is the mass balance:
	\eqn{m_kbal}{
		\tau_b\tau_c\partial_aZ^{abc}  = \dot \rho +\rho \partial_au^a +\partial_a 
		j^\a = 0,
	}
The $u$-spacelike part of \re{mp_kbal} is the momentum balance:
	\eqn{p_kbal}{
		\pi^\b_{\;\;d}\tau_c \partial_aZ^{adc} = \rho \dot u^b + \dot p^\b
		+  p^\b\partial_au^a +  j^\a \partial_au^b	+ \partial_a P^{\a\b} = 0^{\b}.
	}
	
The balance of energy is the $u$-space-spacelike part of the mass-momentum-energy balance, more properly the trace of that:
	\eqn{e_kbal}{
		\f{\delta_{\b\c}}{2} \pi^\b_{\;\;d}\pi^\c_{\;\;e}\partial_a Z^{ade} 
		&= \f{\delta_{\b\c}}{2}\left(
		\dot e^{\b\c}+ e^{\b\c} \partial_au^a + p^\b\dot u^c +   
		p^\c\dot u^b+ P^{\a\b} \partial_a u^c + P^{\a\c} \partial_a u^b  + 
		\partial_aq^{\a\b\c}\right) \nnl{eb2}
		&= \dot e + e\partial_au^a + p^\b\dot u_b +  
		P^{\a}_{\;\;\b} \partial_a u^b +  \partial_aq^\a = 	0.	
	}
	
One may obtain the substantial form of the  balances \re{m_kbal}, \re{p_kbal} and \re{eb2} by using the relative velocity  $v^\a = u^a-u'^a$ of the fluid to the inertial reference frame. With the usual three-index notation they are
	\eqn{sm_bal}{
		\dot \rho+ \rho \partial_i v^i + \underline{\partial_ij^i} &= 0, \nl{sp_bal}
		\underline{\dot p^i + p^k \partial_kv^k} + \rho \dot v^i + 
		\underline{j^k\partial_k v^i} + \partial_kP^{ki} &= 0^i, \nl{e_bal}
		\dot e +e \partial_iv^i +\partial_i q^i + \underline{p_i \dot v^i} + 
		P^{ik}\partial_iv_k  &= 0.
	} 
	
These are the mass, momentum and energy balances of the fluid. The same expressions are obtained if the transformation rules \re{rtr} are applied to exchange the inertial reference frame related $u'$-quantities with medium related $u$-quantities in the local balances. The underlined terms indicate the differences between the usual balances of fluids (see e.g. \cite{Gya70b,Gal02b}) and the ones above. In these terms two additional quantities appear,  the (self)diffusion flux, $j^i$, and the momentum density $p^i$. Therefore the usual closure procedure, prescribing constitutive functions for the pressure tensor, $P^{\a\b}$, and the heat flux, $q^\a$, is insufficient, further conditions are necessary to close the system of equations. To this end one should investigate the thermodynamics of fluids from the point of view of Galilean relativistic space-time.

\section{Thermostatics of motion or thermostatodynamics}

The title of the section reflects the paradoxical dilemma of thermodynamics (or thermostatics?) considering the motion related mechanical 	properties. The literature of thermodynamics rarely introduces velocity as a state variable. In rational continuum mechanical investigations this possibility is forbidden, because the usual theory does not consider classical thermodynamics as a meaningful starting point- Moreover, according to the usual formulation of the principle of material frame indifference  relative velocity cannot be a variable in constitutive functions (see e.g. \cite{JouAta92b} p43). From the point of view of Galilean relativity these statements require further analysis \cite{MatVan06a}.
	
\subsection{Absolute relations.}

Our fundamental point of view is, that classical "equilibrium" thermodynamics is actually time dependent: it is the homogeneous, discrete counterpart of irreversible thermodynamics\footnote{In this respect the book of Tam\'as Matolcsi is clear and instructive \cite{Mat05b}, introducing evolution equations and also the relation of the second law and asymptotic stability of the equilibrium. \cite{Mat05b} settles and extends the issues that were started e.g. by Truesdell and Bharatha \cite{TruBha77b}.}. Moreover, the thermodynamic relations of relativistic kinetic theory can be instructive. The essential aspects are treated in \cite{Van11p,VanBir12a}.
	
The entropy density is a four-vector field, whose $u$-form is $S^a  = s u^a + s^\a$, with the timelike component, the entropy density, $s=\tau_a s^a$ and the $u$-spacelike component, the entropy flux $s^\a = \pi^a_b S^b$. The entropy density is the function of the mass-momentum-energy density: $s = s(z^{bc})$. This is an absolute relation, does not depend neither on a reference frame nor on the velocity field of the medium $u^a$, because both the entropy density and the density tensor of the extensives,  $s$ and $z^{ab}$, are absolute. The derivative of $s$ is the symmetric second order four-cotensor of the thermodynamic intensives, the {\em chemical potential-thermovelocity-temperature four-cotensor}, and denoted by  $\betag_{bc}$ in the following. Therefore, $\f{d s}{d z^{bc}} = \betag_{bc}.$ This derivative is the {\em Gibbs relation} and  in the following it is treated by differentials according to thermodynamic tradition:
	\eqn{aGibbs}{
		ds = \betag_{bc} d z^{bc}.
	}
	
The $u$-form of the chemical potential-thermovelocity-temperature cotensor is:
	\eqn{int_h}{
		\betag_{bc} = \betag_b \tau_c + \betag_{b\eh}\pi_c^{\;\;\eh} = 
		(\bg\tau_b + \bg_\d \pi_b^{\;\;\d})\tau_c +
		(\bg_\eh \tau_b + \bg_{\d\eh} \pi_b^{\;\;\d} )\pi_c^{\;\;\eh}. 
	} 
	
	If the tensor $z^{bc}$ is split to components according to \re{t1}, then
	\begin{itemize}
		\item $y$ is the intensive quantity related to the mass density,
		\item $y_\b$ is the intensive quantity belonging to the momentum density $p^\b$,
		\item $y_{\b\c}$ is the intensive quantity related to the energy density tensor $e^{\b\c}$ .
	\end{itemize}
	
	One can pull closer the treatment to the usual approach assuming
	\eqn{beta}{
		\bg_{\b\c} =\f{\beta}{2}\delta_{\b\c}
	}
	
	In this case
	\eqn{betai}{
		\bg^{\;\;\b}_\b = \delta^{\b\c}\bg_{\b\c}=\f{3}{2}\beta.
	}
	
Then the physical quantities are 
	
	\begin{itemize}
		\item the {\em reciprocal temperature}, $\beta$:
		\eqn{h_def}{
			\beta =\f{1}{T}=  \f{1}{6} \delta^{\b\c} \delta_\c^{\;\;e} 
			\delta_\b^{\;\;d}\betag_{de} =\f{2}{3} \bg^{\;\;\b}_\b.
		}
		\item The {\em chemical potential} $\mu$ is related to the entropic intensive  of the mass density:
		\eqn{kp_def}{
			\mu = -T u^bu^c \betag_{bc} = -T \bg. 
		}
		\item Finally, {\em thermovelocity} is defined with the help of the momentum density related intensive
		\eqn{ts_def}{
			w_\b = -2 T u^c \delta_\b^{\;\;d} \betag_{dc} = -2T \bg_\b.
		}
	\end{itemize}
	
Therefore the $u$-form of the absolute Gibbs relation \re{aGibbs} can be calculated as:
	\eqn{Gibbs_cal}{
		ds &= \betag_{bc}dz^{bc} =  \nnl{Gc1}
		-&\beta\left(\mu \tau_b\tau_c +
		\f{1}{2}(w_\d \pi_b^{\;\;\d}\tau_c + w_\eh \pi_c^{\;\;\eh}\tau_b) 
		-\f{1}{2} \bg_{\d\eh}\pi_b^{\;\;\d}\pi_c^{\;\;\eh} \right) \times\nnl{Gc2}
		& \qquad \left( u^b u^cd\rho +\rho u^c du^b +\rho u^b du^c +
		u^c dp^\b + p^\b du^c+ u^bd p^\c +p^\c du^b+ de^{\b\c} \right)\nnl{Gc3}
		&= -\beta\left( \mu d\rho +\rho w_\b du^b + w_\b dp^\b - p_\b du^b
		- de	\right)
	}
	
	With the help of the  $u$-split quantities finally the following form is obtained:
	\eqn{uGibbs_rel}{
		de = Tds +\mu d\rho +w_\a dp^\a + (\rho w_\a - p_\a)du^a.
	}
	
	This formula is analogous to the relativistic Gibbs relation suggested in \cite{Van11p,VanBir12a,VanBir14p}, where the compatibility with kinetic theory is considered. The enthalpy in the relativistic case is substituted by the mass density here. 
	
	The Legendre transformation of the entropy density four-vector defines the {\em conjugated entropy}, $\tilde S^a$:
	\eqn{ap_def}{
		S^a - \betag_{bc}Z^{abc} = \tilde S^a. 
	} 
	Let us give the $u$-form of the conjugated entropy in the following form
	\eqn{shas}{
		\tilde S^a = \beta p (u^a + r^\a),
	}
	where  $\tau_ar^\a=0$. Then the absolute timelike part of the four-vector equation \re{ap_def}, the {\em extensivity relation} is 
		\eqn{ext_def}{
			s + \beta\mu\rho+\beta w_\b p^\b - \beta e = \beta p. 
		}
	This expression defines the thermostatodynamic (thermostatic) {\em pressure}, $p$.  The $u$-spacelike part, the entropy flux, is obtained with the help of the $u$-projection, $\pi^\a_{\; b}$:
	\eqn{saram_def}{
		s^\a + \beta \mu j^\a +\beta P^{\a\b}w_\b -\beta q^\a =  \beta p r^\a. 
	}
	The extensivity relation \re{ext_def}, and the Gibb relation  \re{uGibbs_rel} together result in the  {\it Gibbs--Duhem relation}:
	\eqn{uGD_rel}{
		\beta dp = -hd\beta + \rho d(\beta\mu )+ p^\a d(\beta w_\a) - \beta (\rho w_\a 
		- p_\a)du^a,
	}
	where $h=e + p$ is the enthalpy density.
	
	\subsection{Equation of state.}
	
The $u$-form of the absolute Gibbs relation, \re{uGibbs_rel} may be interpreted by taking into account that the relative intensives $T, \mu, w_\a, \rho w_\a -p_\a$ are functions of the relative densities $(s,\rho,p^a,u^a)$. Therefore, one of the second order mixed partial derivatives of the internal energy $e$ leads to the following Maxwell-relation:
	\eqn{mpar}{
		 \frac{\partial w_\a}{\partial u^b} =
                      \frac{\partial^2 e}{\partial u^b\partial p^\a} = 
		 \frac{\partial^2 e}{\partial p^\b\partial u^a}	=
                      \frac{\partial (\rho w_\a - p_\a)}{\partial p^\b}.
	}
This is a partial differential equation for $w^\a$, whose general solution is
	\eqn{wsol}{
		w_\a= \frac{p_\a}{\rho} + A_{\a\c}\left(u^c-{\hat u}^c + \frac{p^\c}{\rho}\right),	
	}
	where  $A_{\a\c}(\rho,s)$ and the four velocity ${\hat u}^c(\rho,s)\in V(1)$  are arbitrary functions of $\rho$ and $s$. The above equation of state demonstrates, that the dependence of the thermovelocity on the momentum density $p^\a$ is restricted. The equation of state is very simple if   $A_{\a\c}=0_{\a\c}$. In this case
	\eqn{icon}{
		p_\a=\rho w_\a.
	} 
	In the following we will call this relation {\em momentum condition}.

	\subsection{The transformation rules of thermodynamic relations.}
	
	The entropy density and the entropy flux are time- and spacelike components of the entropy four-vector. The components according to an external inertial observer with constant $u'^a$ four-velocity lead to the corresponding transformation rules between the comoving reference frame to an inertial laboratory one: 
		\eqn{entr}{ 
			s' = s, \qquad  s'^i= s^i + sv^i.
		}
	Here $v^\a = u^a-u'^a$ is the relative velocity field.
	
	The general transformation rules of four-vectors in \re{v_tr} and the representation of $\tilde S^a$ in \re{shas} result in 
	\eqn{ptraf}{
		p' = p \qquad \text{and} \qquad r'^i = r^i + v^i.
	}
	
	The components of the absolute chemical potential-thermovelocity-temperature four-cotensor, $Y_{ab}$,
	are transformed acording to \re{kk_tr} of Appendix B:
	\eqn{btr}{
		\beta' &= \beta, \nl{wtr}
		w'^i  &= w^i + v^i, \nl{mutr} 
		\mu'  &= \mu - w_iv^i - \f{v^2}{2}.
	}
	
	It is worth to analyze thermovelocity with more details. The spacelike part of a second order cotensor is absolute: $y_{\a b}= \delta_\a^{\;\;c}Y_{cb} = y_\a \tau_b + y_{\a\c}\pi_b^{\;\;c}$. Its $u'$-timelike component is 
	\eqn{trtermo1}{
		y'_\a = y_{\a b} u'^b = y_\a + y_{\a\c} (u'^c - u^c),
	} 
	Therefore, for the thermovelocity one obtains with the help of \re{beta} and \re{ts_def}:
	\eqn{tstra}{
		w'_\a = w_\a - \delta_{\a\c} (u'^c - u^c) = w_\a + v_{\a},
	}
	Written with relative indexes this is the transformation rule \re{wtr}.
	
	The extensivity relation is absolute and therefore Galilean invariant, because it is the absolute timelike part of an four-vector equation:
	\eqn{ext_rel}{
		e'+p - T s - \mu' \rho - w'_i p'^i = e+p - T s - \mu \rho - w_i p^i = 0. 
	}
That can be verified also directly with the derived particular transformation rules of the physical quantities. A similar calculation shows, that the entropy flux transforms as a space vector.
	\eqn{sá_traf}{
		s'^i = \beta (q'^i - \mu' j'^i - P'^{ij}w'_j + p r'^i) = \beta (q^i - \mu j^i - 
		P^{ij}w_j + p r^i) + s v^i = s^i + s v^i,
	}
	where the previously derived transformation rules of $\beta$, $q^i$, $\mu$, $j^i$, $P^{ij}$, $w_i$, $p$ és $r^i$ were applied. 
	
	Finally, the Gibbs relation \re{uGibbs_rel} is Galilean invariant, too, because it is obtained applying a cotensor to a tensordifferential, 
	\eqn{Gibbs'_rel}{
		de- Tds - \mu d\rho - w_idp^i - (\rho w_i - p_i)dv^i=de'- Tds - \mu'd\rho - w'_idp'^i - (\rho w'_i - p'_i)d{\hat v}^i,
	} 
	where $v^\a = u^a-u'^a$ and ${\hat v}^\a = u^a-u''^a$ are the relative velocities of the medium with respect to inertial observers with constant $u'$ and $u''$ four-velocities. The last term of the Gibbs relation is Galilean invariant in itself, like the $Tds$ term, because $\rho w_i - p_i = \rho w'_i - p'_i$ and $dv^i=d{\hat v}^i$.

	\section{Absolute entropy production}

	With the $u$-forms of the Gibbs relation  and the entropy flux, \re{uGibbs_rel} and \re{saram_def}, the entropy balance can be expressed as:
	\eqn{as_bal}{
		\partial_aS^a &=  \dot s +s \partial_a u^a + \partial_as^\a = \nnl{sb1}
		&\quad \beta\dot e - \beta\mu \dot \rho -\beta w_\a \dot p^\a 
		+\beta(p_\a-\rho w_\a)\dot u^a +s \partial_au^a + \nnl{sb11}
		&\qquad
		\partial_a\left(\beta q^\a -\beta\mu j^\a -\beta P^{\a\b}w_\b + \beta p 
		r^\a  \right).
	}
	Substituting the balances of mass, momentum and energy, \re{m_kbal}, \re{p_kbal} and \re{eb2}, then using also the Gibbs-Duhem relation \re{uGD_rel}, one obtains, that
	\eqn{as2_bal}{
		\partial_aS^a &=  (s-\beta e+  \beta \mu \rho +\beta w_\a p^\a )\partial_au^a + 
		q^\a \partial_a \beta -j^\a\partial_a (\beta\mu ) - \nnl{sb3}
		&\beta P^{\a}_{\;\;\b} 
		\partial_a(u^b+w_\b) + \beta w_\a j^\b \partial_b u^a - P^{\a\b} w_\b 
		\partial_\a\beta +\beta p \partial_a r^\a + r^\a \partial _a(\beta p) =  
		\nnl{sb4}
		& \left(\rho r^\a - j^\a\right)
		\left(\partial_a (\beta\mu) - \beta w_\b \partial_a u^b \ \right) + 
		\nnl{sb41}
		&\qquad\left(q^\a -h r^\a -(P^{\a\b}-r^\a p^\b )w_\b + pr^a  \right)
		\partial_a\beta -
		\nnl{sb42}
		&\qquad	\beta\left(P^\a_\b- r^\a p_\b -p\delta^\a_{\;\;\b}\right)
		\partial_a(u^b+w^\b) +
		\beta p	\partial_a(r^\a-w^\a) = \nnl{sb5}
		& \left(\rho r^\a - j^\a\right)
		\partial_a \left(\beta\left(\mu + \f{w^2}{2} \right)\right) +
		\nnl{sb51}
		&\qquad	\left(q^\a -e r^\a -(P^{\a\b}-r^\a p^\b )w_\b -
		(\rho r^\a - j^\a) \f{w^2}{2}\right)\partial_a\beta -
		\nnl{sb52}
		&\qquad	\beta\left(P^\a_{\;\;\b} - j^\a w_\b - r^\a (p_\b-\rho w_\b) -
		p\delta^\a_{\;\;\b} \right)\partial(u^b+w^\b)+
		\beta p\partial_a(r^\a-w^\a)\geq 0.
	}
	
	This inequality of the absolute entropy production is the second law of Galilean relativistic single component fluids. This is a quadratic expression. The first term expresses the diffusion related dissipation, the constitutive quantity is the 	(self)-diffusion flux, $j^\a$. In the second part of the product there is the gradient of the  chemical potential divided by the temperature. The second term is the thermal part of the dissipation, where the constitutive quantity is the heat flux,  $q^\a$. The gradient of the reciprocal temperature, $\partial_a\beta$ is the thermodynamic force. The third term is related to mechanical dissipation, with the pressure tensor, $P^{\a\b}$, as constitutive quantity, and with the gradient of a velocity as thermodynamic force. The relevant velocity is the sum of the  $u^a$ four-velocity and the $w^\a$ thermovelocity. Let us observe, that  $u^a$ appears explicitely only here, in the velocity gradient part of the  dissipation. The fourth term is new. Here $r^\a$ can be the constitutive quantity. 

	Therefore the inequality of the absolute entropy production seems to be solvable in the sense that in every term there is a constitutive quantity, therefore one can introduce thermodynamic fluxes and forces and assume a linear relationship with positive definite coupling between them. However, the thermodynamic conditions alone do not close the system of balances. Enumerating the variables one can conclude that there is no differential equation either for the momentum  or for the velocity of the continuum. The thermovelocity should be

	determined or fixed, too. Let us recognize here, that up to know we did give any physical condition that would connect $u^a$ to the medium, therefore it is not known  yet the physical meaning of the it is not known  yet the physical meaning of the velocity field of the fluid.

\section{What is the velocity of a fluid?}

For the sake of simplicity in the following we do not investigate the last term of the entropy production, therefore the spacelike part of the conjugated entropy is considered parallel to the thermovelocity, as it is customary in special relativistic kinetic theory. In particular our assumption is, that
\eqn{pdef}{
		r^\a = w^\a.
	}
	
Then the entropy production simplifies to the following form:
	\eqn{sp_tr}{
		\partial_aS^a &= \left(\rho w^\a - j^\a\right)
		\partial_a \left(\f{\mu}{T}+ \f{w^2}{2T} \right) +\nnl{spt1}
		&\quad	\left(q^\a -w^\a(e-p^\b w_\b) -(\rho w^\a - j^\a )\f{w^2}{2} -
		P^{\a\b}w_\b \right)\partial_a\f{1}{T} - \nnl{spt2}
		&\quad \f{1}{T}\left(P^\a_{\;\;\b} - j^\a w_\b - w^\a (p_\b-\rho w_\b) -
		p\delta^\a_{\;\;\b} \right)\partial_a(u^b+w^\b)\geq 0.
	}

How could we reduce the number of unknowns to obtain a closed system of equations? In addition to the usual system of basic variables, the density, the internal energy density and the velocity field, $\rho, u^a, e$, we have two other fields: the momentum density and the thermovelocity,
$p^\a$ and $w^\a$. We have already discussed the equation of state for the thermovelocity. On the other hand up to know we did not fix what is the meaning of the four-velocity of the fluid. This choice, the physical definition of the fluid velocity is the {\em flow-frame}.  
	
There are several possibilities. We may fix the velocity field $u^a$ to one of the extensive quantities of the fluids, e.g. to the mass ($j^\a=0$), to the energy ($q^\a=0$) and also to the momentum density ($p^\a=0$), but it can be fixed by other, more complicated ways, too. With the previous definitions relative velocity is the flow of the mass, energy or momentum related to an external observer. An example of more complicated choices is mixture of energy- and particle(mass)-flow \cite{VanBir14p}. The different choices are not equivalent from a practical point of view. Looking at the above expression of the entropy production one may recognize, that a simple form of constitutive functions is obtained fixing the four velocity of the fluid to the thermovelocity, that is $w^\a=0$. This flow-frame is called {\em thermo-flow}.

The entropy production with thermo-flow is the following:
	\eqn{sp_wrel}{
		\partial_aS^a = 
		- j^\a \partial_a \f{\mu}{T}+ 
		q^\a\partial_a\f{1}{T} - 
		\f{1}{T}\left(P^\a_{\;\;\b} -p\delta^\a_{\;\;\b}\right)\partial_a u^b \geq 0.
	}
	
	
If the momentum condition equation of state, \re{icon}, is applied, then $w^\a = 0^\a$ is the consequence of $p^\a = 0^\a$. Therefore in this case a thermo-flow is necessarily a {momentum-flow}, too.
	
A fluid, defined with thermo- and momentum-flows is called {\em classical fluid}. Then the substantial mass-, momentum-, and energy balances become simpler, too: 
	\eqn{sm_fbal}{
		\dot \rho+ \rho \partial_i v^i + {\partial_ij^i} &= 0, \nl{sp_fbal}
		\rho \dot v^i + {j^k\partial_k v^i} + \partial_kP^{ki} &= 0^i, \nl{e_fbal}
		\dot e +e \partial_iv^i +\partial_i q^i +  P^{ik}\partial_iv_k  &= 0,
	} 
and the entropy production is
	\eqn{sp_srel}{
		\partial_aS^a = 
		- j^i \partial_i \f{\mu}{T}+  
		q^i\partial_i\f{1}{T} - 
		\f{1}{T}\left(P^{ij} -p\delta^{ij}\right)\partial_i v_j \geq 0.
	}
	
For classical fluids the velocity dependent Gibb-relation, \re{uGibbs_rel}, reduces to the usual form
	\eqn{klGibbs}{
		de = Tds + \mu d\rho.
	}

In case of inertial observers $w'_i=v_i$ and  $p'^i=\rho v^i$, therefore $\rho w'^i - p'^i=0$. As a consequence, the transformed form of the Gibbs relation, \re{Gibbs'_rel}, becomes
	\eqn{Gibbs_frel}{
		de' = Tds +\mu' d\rho + v_id(\rho v^i).
	}
Similarly the extensivity relation, \re{ext_def},
	\eqn{extT_rel}{
		e'+p = Ts +\mu'\rho + \rho v^2 
	}
can be written in the form
	\eqn{extb_rel}{
		e+p = Ts +\mu\rho.
	}
These formulas show well, that for classical fluids, the so called total energy density is the  $u'$-energy density $e'$ corresponding to  the inertial reference frame, and the $u$-energy density $e$ is the internal energy. They are related by the transformation rule \re{etr0}. Similarly, in the transformation rule of chemical potential, $\mu' = \mu- \frac{v^2}{2}$, we could call $\mu'$ as total and $\mu$ as internal chemical potential. In the customary extensivity relation the term of the kinetic energy is merged into pressure, instead of chemical potential. In fluid mechanics the extensivity relation appears as Bernoulli equation with the so called dynamic pressure.
	
The $u'$-form of the entropy flux of classical fluids provides the  {\em total entropy flux}:
	\eqn{saramf}{
		s'^i = \f{1}{T} \left(q'^i -\mu' j'^i - P'^{ik}v_k + p v^i\right),
	}
	and the $u$-form is the {\em internal entropy flux}:
	\eqn{saramfb}{
		s^i = \f{1}{T} \left(q^i -\mu j^i\right),
	}
	
	The entropy production, \re{sp_srel}, can be also calculated with the help of the relative balances and thermodynamic relations, \re{sm_fbal}-\re{e_fbal}, \re{klGibbs}, \re{extb_rel} and \re{saramfb}. 
	
	The corresponding thermodynamic forces and fluxes are given in table \ref{erar}.
	{\center
		\begin{table}
			\begin{tabular}{l c c c}
				& Diffusion & Thermal & Mechanical\\ \hline\\
				Force & -$\partial_i\frac{\mu}{T}$ & $\partial_i\frac{1}{T}$ & $\partial_iv_j$\\ 
				Flux & $j^i$ & $q^i$ &	-$\frac{1}{T}\left(P^{ij}- p\delta^{ij}\right)$
			\end{tabular}
			\label{erar}\caption{Thermodynamic forces and fluxes}
		\end{table}}
	According to the representation theorems of isotropic functions \cite{Mul85b}, for isotropic fluids the linear relationship between the thermodynamic fluxes and forces results in the following constitutive functions
		\eqn{diff_c}{
			j^i &= -\xi \partial_i\f{\mu}{T} + \chi_1 \partial_i\frac{1}{T},\nl{hvez_c}
			q^i &= -\chi_2 \partial_i\f{\mu}{T} + \lambda \partial_i\frac{1}{T},\nl{viszk_c}
			P^{ij} &= p\delta^{ij} - \eta_v \partial_k v^k \delta^{ij} -
			\eta \left(\partial^iv^k+\partial_kv^i - \f{2}{3} \partial_k v^k 
			\delta^{ij} \right).
		}
		
		Here $\xi$ is the (self)diffusion coefficient, $\chi_1$ and $\chi_2$ are the (self)Soret--Dufour coefficients, $\lambda$  is the thermodynamic coefficient of heat conduction ($\lambda_F = T^2 \lambda$ is the  Fourier heat conduction coefficient), $\eta_v$ and $\eta$ are the volume and shear viscosities.
		
		The \re{sm_fbal}-\re{e_fbal} system of basic balances, together with \re{diff_c}, \re{hvez_c} and the \re{viszk_c} constitutive functions is a closed system of equations, and with the notable exception of the (self)diffusion flux, it is identical with the usual continuity-Fourier-Navier-Stokes system of equations. Our derivation shows, that the (self)diffusion flux cannot be eliminated by flow-frame choice, its neglection is a physical assumption about material properties.

		\section{Summary}
		
		Non-relativistic, more properly, Galilean relativistic single component dissipative fluids were treated independently of reference frames and flow-frames. The particular conditions leading to the usual relative, reference frame dependent continuity-Fourier-Navier-Stokes system of equations were given. The reference frame free theory is based on the Galilean relativistic space-time model of Matolcsi \cite{Mat86b,Mat93b,Mat15b}. Our treatment here uses mostly vector spaces and introduces an adapted abstract index formalism. 
		
		It was required, that usual relative physical quantities must be components of absolute ones, and that the tensorial properties should be in harmony with the momentum series expansion of the kinetic theory. The analysis of these requirements leads to our basic physical quantity, to a third order, partially symmetric four-tensor, called the mass-momentum-energy density-flux tensor of the fluid. The four-divergence of this quantity is a second order four-tensor differential equation representing unitedly the reference frame independent form of the mass, momentum and energy balances. With the help of a four-velocity field the mass balance emerges as the time-timelike component, the momentum balance as the time-spacelike component and also as the space-timelike component according to the symmetry. The energy balance is the trace of the space-spacelike component of the absolute balance. 
		
		The Galilean transformation rules of the particular relative physical quantities and balances were derived. One of the consequences of the theory is that the usual relation between the internal, kinetic and total energies appears as a transformation rule. The derived  transformation rule of the energy density, \re{etr}, is more general than the usual expression with internal, total and kinetic energies, \re{etr0}, because of the presence of (self)momentum density. The transformation rules of the physical quantities are transitive.
		
		Regarding thermodynamics our basic assumption was, that the entropy density depends on the mass-momentum-energy density.  This is an absolute statement, independent of reference frames and flow-frames. The derivative of the entropy density provides the second order four-cotensor of intensive quantities, the absolute temperature-thermovelocity-chemical potential cotensor. The intensive pair of the momentum density, the thermovelocity, appears in the $u$-form of the Gibbs relation, too. The equations of states of the relative intensives are not independent of each other, due to the absolute background. The equality of second order mixed partial derivatives of the motion related intensive quantities, the corresponding Maxwell relation, can be solved and restricts the form of the thermovelocity equation of state. 
		
		The particular relative form of the entropy flux follows from the basic assumptions, therefore the four-divergence of the entropy four-vector, that is the absolute entropy production, can be calculated. There is a freedom to fix the flow-frame of a fluid to the mass (Eckart flow-frame), to the energy (Landau-Lifsic flow-frame) or by other, different manners.
		
		The particular form of the entropy production reveals, that the most convenient choice of the flow-frame is to fix it to the temperature, and eliminate the thermovelocity. This is thermo-flow. Therefore the momentum density is zero, too, because of the thermovelocity equation of state. With this condition we almost obtain the usual form of the entropy production. The deviation comes from the presence of the (self)diffusion terms. (Self)diffusion cannot be eliminated any more by changing the flow frame if we want to keep the usual form of the other terms. 

Remarks: 
		
	\begin{itemize}
		\item Representing the basic field of the continuum in a Galilean relativistic space-time by a third order mixed tensor field, instead of third order tensor field, we obtain the same relative balances with the same transformation rules. Then the second order spatial component of the density four-tensor will be the mass density, instead of the energy. However, the covariant parts of the mixed tensor are not compatible with the usual system of equations obtained from momentum series expansion in kinetic theory.
			
		\item The system of equations \re{sm_fbal}-\re{e_fbal} and \re{diff_c}-\re{viszk_c} is different than the most investigated similar system of Brenner (see e.g. \cite{BedEta06a}). Therefore the arguments against the presence of this term should be reconsidered \cite{VanEta15m}.
			
		\item The stability of simple materials under simple environmental conditions is an essential experimental observation in physics. Without this fundamental stability property physical phenomena may not be reproduced \cite{PriStr95b}. The physical-mathematical representation of this stability is thermodynamics. The concepts  and the structure of thermodynamics can be understood from this point of view \cite{Mat92a1,Van94p,Van95a,Mat05b}. This approach to thermodynamics is also a benchmark, a tool of verification of a thermodynamic theory. In case of Galilean relativistic fluids we expect, that the homogeneous equilibrium be asymptotically stable without any additional conditions beyond the thermodynamic requirements \cite{VanBir08a,Van09a}. For classical fluids the following statement can be proved:
			
			The system of equations \re{sm_fbal}-\re{e_fbal} and \re{diff_c}-\re{viszk_c} is generic stable, that is its homogeneous equilibrium is linearly stable, if thermodynamic stability is fulfilled (entropy density is concave), the transport coefficients are nonnegative (second law) and the following inequality is true:
			\eqn{sfelt}{\xi \f{\partial }{\partial \rho} \f{\mu}{T} - 
				\lambda \f{\partial }{\partial e} \f{1}{T} +
				(\chi_1+\chi_2)  \f{\partial }{\partial e} \f{\mu}{T} \geq 0.
			} 
			Here the first two terms and the coefficient of the third term is nonnegative because of thermodynamic conditions, but the partial derivative of the last term may lead to the violation of the inequality.
			
			\item The main motivation of this work comes from our attempts to clarify the relation of generic stability, thermodynamics and flow-frames in dissipative relativistic fluids: \cite{VanBir08a,Van08a,BirAta08a,Van09a,VanBir09dem,BirVan10a,Van11p,VanBir12a,VanBir13p,VanBir14p}. 
		\end{itemize}

\section{Appendix A. Galilean relativistic space-time}
\label{App_alap}

The mathematical structure of the Galilean relativistic space-time is 

\begin{enumerate} 
\item $M$ {\em space-time} is an oriented four dimensional affine space of the {\em world points} {\em or events}  $x\in M$, over the four dimensional vector space of {\em space-time vectors} $x^a\in \Mi$. There are no Euclidean of pseudo-Euclidean  structures on $\Mi$: the length of a space-time vector does not exist.
\item $I$ {\em time} is a one dimensional oriented affine space of {\em instants}  $\text{t} \in I$, over the one dimensional vector space of {\em durations}  $t\in \I$. 
\item $\mathbb{\tau}:  M \rightarrow I$ is the {\em timing}, an affine surjection over the linear mapping  $\tau_a: \Mi \rightarrow \I$, the {\em time evaluation}. 
\item $\D$ is the {\em  measure line of  distances}, which is a one dimensional oriented affine space.
\item $\delta_{\a\b}:\E \times \E\rightarrow \D\otimes \D$ Euclidean structure is a symmetric bilinear mapping, where $\E := Ker(\tau_a) \subset \Mi$ is the three dimensional vector space of {\em space vectors}. 
\end{enumerate}

The empirical--axiomatic foundation of this Galilean relativistic space-time is given in \cite{Mat15b}, where the axioms are related to clear observations and measurements. This structure was given first in \cite{Mat84b} and further elaborated in \cite{Mat93b}. Similar structures where suggested also in \cite{Fri83b,CarCha04a}.


The {\em duration} between the events $x,y \in M$ is calculated by $\tau(x)-\tau(y) = \tau_a x^a$, where $x^a = x-y$. Two events are {\em simultaneous} if the duration between them is zero. The difference between two simultaneous events is a {\em spacelike} vector, an element of $\E$. Those vectors that are not spacelike are called {\em timelike}.

The dual of $\Mi$, the vector space of the  $\Mi\rightarrow \mathbb R$ linear mappings, is denoted by $\Mi^*$. the elements of  $\Mi^*$ are called four-covectors and are denoted by lower indexes. Similarly, the dual of $\E$ is $\E^*$, and their elements, the spacelike vectors and covectors, are denoted by overlined upper or lower indexes $x^\a\in\E$, $x_\a\in\E^*$, respectively. The {\em length} of a spacelike vector is $\|x\| = \sqrt{x^\a \delta_{\a\b}x^\b}$.

There is a canonical identification of $\E$ and $\E^*$, due to the Euclidean structure. However, $\Mi$ and $\Mi^*$ cannot be identified for lack of  Euclidean or pseudo-Euclidean structures: timelike space-time vectors do not have a length. 

\begin{figure}
\centering
\includegraphics[scale=0.4]{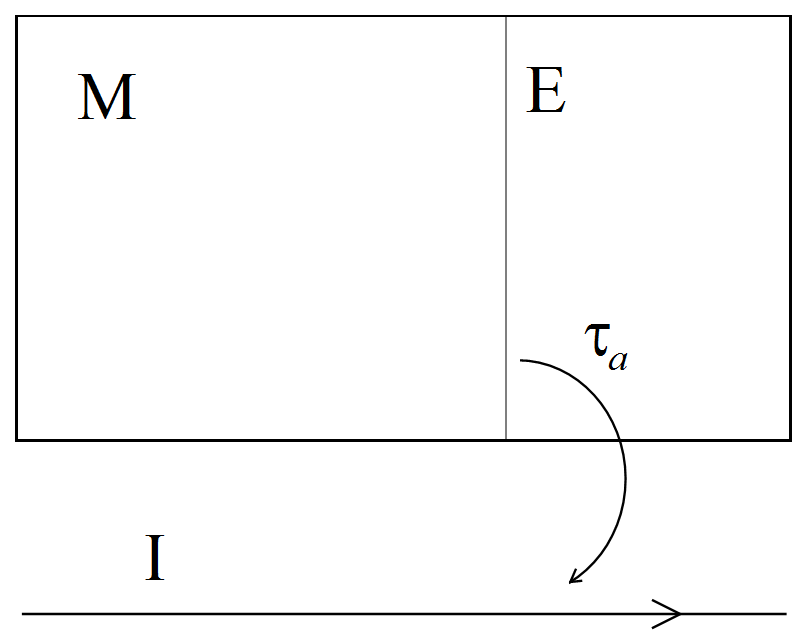}
\caption{ \label{st_fig}
 The relation of Galilean relativistic space, time and space-time.}
\end{figure}

The most important elements of the model are shown on figure \ref{st_fig}. Time evaluation and timing introduce a foliation of space-time: the sequence of spacelike subspaces of simultaneous events. 

We use an abstract index notation of vectors and covectors. $a,b,c,d,e,f,g$ indexes are used for absolute physical quantities in the four-dimensional space-time. The indexes are abstract in the sense, that do not refer to a particular coordinate system or reference frame, they denote the tensorial properties of the different space-time related physical quantities \cite{Pen07b}. 
Upper indexes are used for vectorial (contravariant), the lower indexes for covectorial (covariant) physical quantities. The $i,j,k,l,m$ indexes always refer to tensorial components of usual three dimensional relative physical quantities as seen in an inertial reference frame. These indexes indicate the presence of two absolute velocity fields, here the velocity fields of the inertial observer and that of the fluid, and can be rewritten with absolute indexes, where the corresponding four-velocities are explicitly written. We will work with a space-time notation, but for the interpretation of the traditional forms of the equations and the transformation formulas of Galilean transformations the later indexes are used.

The vector space $\E$ of spacelike vectors is a subspace of $\Mi$, and its canonical embedding is denoted by $\delta^a_{\;\b}$. Therefore if $x^\a\in\E$, then  $\delta^a_{\;\b}x^\b\in\Mi$ is nothing else but $x^\a$ as the element of $\Mi$.

If $x_a$ is a covector, that is a linear mapping $x_a:\Mi\rightarrow \mathbb{R}$, then its restriction to $\E$ is an element of $\E^*$ denoted by $x_\a$. The corresponding {\em projection of the restriction} is denoted by $\delta^{\;a}_{\b}\in Lin(\Mi^*,\E^*)$, therefore $\delta^{\;b}_{\a} x_b = x_\a$. 

This projection $\delta^{\;a}_{\b}\in Lin(\Mi^*,\E^*)$ is the dual (transpose) of the canonical embedding $\delta^a_{\;\b}\in Lin(\E,\Mi)$. The identification of $\E$ and $\E^*$ by the Euclidean structure is given by $\delta_{\a\b}$ and its inverse  $\delta^{\a\b}$, as  $x_\a=\delta_{\a\b}x^\b$ and $x^\a=\delta^{\a\b}x_\b$. However, one cannot introduce a canonical identification of  $\Mi$ and $\Mi^*$, because of the absence of an Euclidean or pseudo-Euclidean structure.

The above system of notations of vectors and covectors introduces a convenient formal tool of handling the space-time without embedded time. 

\subsection{Splittings}

The existence of a mass point in space-time is given by {\em world line functions}, that map time into space-time $r: I \rightarrow M$. The structure of Galilean relativistic space-time requires, that $\tau(r(t)) = t$. Therefore the time derivative of a world line function at an instant is a four-vector $u^a$ with the following property: $\tau_a u^a =1$. The projection  to $\E$ along an $u^a$ of gives the $u$-spacelike component of a vectorial physical quantity. This {\em $u$-projection } is  $\pi(u)^\a_{\; b} =\delta^a_{\; b} - u^a\tau_b : \Mi \rightarrow \E$, where $\delta^a_{\; b}$ is the identity of $\Mi$.
 
The four-velocities may play an other role: they can map duration into space-time vector  $u^a: \I \rightarrow \Mi, t\mapsto u^a t$. 

Let us enumerate the four basic mappings of space-time vectors:
\begin{itemize}
\item $\tau_a: \Mi\rightarrow \I$,
\item $u^a: \I\rightarrow \Mi$,
\item $\pi(u)^\a_{\; b} = \delta^a_{\; b} - u^a\tau_b: \Mi\rightarrow \E$,
\item $\delta^a_{\; \b}:  \E\rightarrow \Mi$,
\end{itemize}

The corresponding mappings between dual spaces are:

\begin{itemize}
\item $\tau_a: \I^*\rightarrow \Mi^*$,
\item $u^a: \Mi^*\rightarrow \I^*$,
\item $\pi(u)^{\;\b}_{a}: \E^*\rightarrow \Mi^*$,
\item $\delta^{\;a}_{\b}:  \Mi^*\rightarrow \E^*$.
\end{itemize}

The next identities follow from the definition
\eqn{indazon}{
\tau_a u^a =1, \quad
\tau_a \delta^a_{\; \b} = 0_\b, \quad
\pi(u)^\b_{\;a}u^a =(\delta^b_{\;a}- u^b\tau_a)u^a =  0^\b\quad
\pi(u)^\a_{\;b}\delta^b_{\;\c} = \delta^{\a}_{\;\c}.
} 

One can see also that $\delta_\c^{\;a}\pi_a^{\;\b}(u)=\delta_\c^{\;\b}$, and  $\tau_au^c + \pi_a^{\;\b}(u)\delta_{\b}^{\;c}= \delta_a^{\;c}$, therefore $\pi(u)_a^{\;\b}\delta_{\b}^{\;c}\neq \delta_a^{\;c}$. Let us observe, that from the second and last equalities of \re{indazon} result in $\tau_ax^{\a}=0$ and $\pi(u)^\a_{\;b}x^{\b}=x^\a$. 

These relations are summarized by the following diagrams:
$$ \displaystyle
\E \begin{array}{c}
\overset{\pi(u)^\a_{\; b}}{\longleftarrow} \\[-8pt]
\underset{\delta^a_{\; \b}}\longrightarrow
\end{array} \Mi 
\begin{array}{c}
\overset{\tau_a}{\longrightarrow} \\[-8pt]
\underset{u^a}{\longleftarrow}
\end{array} \I, \qquad\qquad
\E^*\begin{array}{c}
\overset{\delta^{\;a}_{\b}}{\longleftarrow} \\[-8pt]
\underset{\pi(u)^{\;\a}_{b}}\longrightarrow
\end{array} \Mi^*
\begin{array}{c}
\overset{u^a}{\longrightarrow} \\[-8pt]
\underset{\tau_a}{\longleftarrow}
\end{array}\I^*.
$$

The upper lines of the above diagrams give the splittings of a vector and a covector into time- and spacelike parts.

\subsection{$u$-form of a vector}
A four-vector can be split into components with respect to an $u^a$ four-velocity and can be reconstructed with the help of these $u$-components.  If $A^a$ is a four-vector,  its timelike part is $A=\tau_aA^a$ and its $u$-spacelike part is $A(u)^\a = \pi(u)^\a_{\;b}A^b$. The timelike part of a vector does not depend on $u$, therefore it is absolute. The spacelike part depends on the four-velocity of the splitting. The reconstruction is simply
\eqn{vek_uf}{
	A^a = Au^a+A(u)^\a.
}
This formula will be called the {$u$-form of the vector}.

The timelike part of an arbitrary four-velocity is $1$. The spacelike part of $u^a$ by an other velocity $u'^a$, 
$$
\pi(u')^\a_{\;\;\;b}u^b = (\delta^a_{\; b} - u'^a\tau_b)u^b= u^a-u'^a = v^\a,
$$
is the {\em relative velocity} of $u^a$ related to $u'^a$. In particular the relative velocity of $u^a$ with respect to itself is zero. 

The extensive quantities together with their fluxes are natural four-vectors in Galilean relativistic space-time. The densities of extensive quantities are four-vector fields. The timelike part of an extensive four-vector density is the  density, an $u$-spacelike part is the flux. The timelike parts of the four-densities are independent of any splitting velocity, their $u$-spacelike parts are not. We will see, that $u$-independency means Galilean invariance.

\subsection{$u$-form of a covector}
Covectors can be split into $u$-timelike and spacelike parts and can be reconstructed with the help of these $u$-components. A covector $B_a$ can be written as
\eqn{kovek_uf}{
B_a = B(u)\tau_a + \pi(u)_a^{\;\b}B_\b.
}
where $B=u^aB_a$ and $B_\a = \delta_\a^{\;b}B_b$. 

Here $B_\a$, the spacelike part of $B_a$, appears differently than the spacelike part of vectors in \re{vek_uf}. One should pay attention that $\pi(u)_a^{\;\b}$ cannot be decomposed additively, more properly  $u^b B_\b$ is not meaningful, because $B_\b \in \E^*$ and $\E^*$ is not a subset of $\Mi^*$. Therefore the convenient regrouping  $B_a = B(u)\tau_a + B_\a - \tau_au^\b B_\b = (B(u)-u^bB_\b)\tau_a +B_\a$ of the above formula is strictly speaking incorrect. However, the spacelike + timelike composition is very transparent and also helpful in the calculations. The advantage of transparent calculations is larger than the possibility of mistakes, therefore we will use this convenient decomposition, with and extra care for the presence of both parts of $\pi(u)_a^{\;\b}$ in the formulas.

The space-time derivative, $\partial_a$, is a covector. It can be written with its $u$-timelike and spacelike components as 
\eqn{der_urep}{
	\partial_a =\tau_a D_u +  \pi(u)_a^{\;\b} \nabla_\b =(D_u- u^{\;b} \nabla_\b)  
	\tau_a+  \nabla_\a,
}
where $D_u = u^a \partial_a$ is the $u$-timelike derivative and $\nabla_\a$ is the spacelike derivative. The spacelike derivative is absolute.

\subsection{$u$-form of a tensor}

The $u$-form of the second order tensor $T^{ab} \in \Mi\otimes\Mi$ is the following
\eqn{kkten_uf}{
T^{ab} = t^a u^b +t^{a\b}= 
 u^a t^b +t^{\a b} = 
 t u^a u^b+ u^a t^\b+t^\a u^b+t^{\a\b},
}
where
\begin{itemize}
\item $t=\tau_a\tau_b T^{ab}$ is the {\em time-timelike} part of $T^{ab}$. It is absolute, independent of $u^a$.
\item $t^\a = \pi(u)^\a_{\;b}T^{bc}\tau_c$ is the {\em space-timelike} part of $T^{ab}$, and $t^\b = \tau_cT^{ca}\pi(u)^{\;\b}_{a}$ is the  {\em time-spacelike} part.
\item $t^{\a\b} = \pi(u)^\a_{\;c}T^{cd}\pi(u)^{\;\b}_{d}$  is the  {\em space-spacelike} part of the tensor.
\item  $t^a = \tau_b T^{ab}$ and $t^b = \tau_a T^{ab}$. The tensor $T^{ab}$ itself is independent of $u^a$, therefore its left and right timelike parts are absolute. If $T^{ab}$ is symmetric, then $\tau_b T^{ab} =\tau_b T^{ba}= t^a$. 
\item $t^{a\b} = \pi(u)^\b_{\;c}T^{ac}$ and $t^{\a b} = \pi(u)^\a_{\;c}T^{cb}$ are the left and right spacelike parts of $T^{ab}$.
\end{itemize}

\subsection{$u$-form of a mixed tensor}

The $u$-form of the second order mixed tensor $Q^{a}_{\;b} \in \Mi\otimes\Mi^*$ is:
\eqn{kkoten_uf}{
Q^{a}_{\;b} =
q^a\tau_b + \pi(u)_b^{\;\c}q^a_{\;\c} =  
q u^a \tau_b+ q^\a\tau_b + u^a \pi(u)_b^{\;\c}q_\c+q^\a_{\;\c} 
\pi(u)_{b}^{\;\c}= \nnl{s1}
\left(u^a(q-u^cq_\c)+q^\a-q^\a_{\;\c}u^c \right)\tau_b + q_\b u^a + q^\a_{\;\b},
}
where
\begin{itemize}
\item $q=u^b\tau_aQ^a_{\;b}$, is the {\em time-timelike part} of $Q^{a}_{\;b}$,
\item  $q^\a = u^b\pi(u)^\a_{\;c}Q^c_{\;b}$, is the {\em space-timelike part} of $Q^{a}_{\;b}$, 
\item  $q_\b = \tau_a\delta_\b^{\;c}Q^a_{\;c}$, is the {\em time-spacelike part} of the mixed $Q^{a}_{\;b}$ tensor. This part is $u$-independent, therefore absolute,
\item  $q^\a_{\;\b} = \pi(u)^\a_{\;c}\delta_\b^{\;d}Q^c_{\;d}$, is the {\em space-spacelike part} of $Q^{a}_{\;b}$,
\item $q^a=u^bQ^a_{\;b}$, is the {\em co-timelike part} of $Q^{a}_{\;b}$,
\item $q^a_{\;\b}= \delta^c_{\;\b}Q^a_{\;c}$ is the {\em co-spacelike part} part of the mixed tensor. It is absolute.
\end{itemize}

The symmetry of a mixed tensor cannot be interpreted $u$-independently.

\subsection{$u$-form of a cotensor}

The $u$-form of the second order cotensor $R_{ab} \in \Mi^*\otimes\Mi^*$ is:
\eqn{kokoten_uf}{
R_{ab} &=
r_a\tau_b + r_{a\c}\pi(u)_b^{\;\c}  = 
\tau_a r_b + r_{\c b}\pi(u)_a^{\;\c}  \nnl{kkt_uf0}
&= r \tau_a\tau_b + r_\c\pi(u)_a^{\;\c}\tau_b + r_\c\tau_a\pi(u)_b^{\;\c} +
r_{\c\d}\pi(u)_b^{\;\c} \pi(u)_{a}^{\;\d} 
\nnl{kkt_uf}
&=\left(r-2r_\c u^c+r_{\c\d}u^cu^d \right)\tau_a\tau_b + 
\left(r_\b-r_{\b\d}u^d\right)\tau_a + \left(r_\a-r_{\a\d}u^d\right)\tau_b +
r_{\a\b},
}
where
\begin{itemize}
\item $r=u^au^bR_{ab}$, is the {\em time-timelike part} of the $R_{ab}$ cotensor,
\item  $r_\a = \delta_\a^{\;c} u^bR_{cb}$, is the {\em space-timelike part} of $R_{ab}$, and  $r_\b = \delta_\b^{\;c}u^aR_{ac}$, is the {\em time-spacelike part},
\item  $r_{\a\b} = \delta_\a^{\;c}\delta_\b^{\;d}R_{cd}$, is the {\em space-spacelike part} of the $R_{ab}$ cotensor. This is the $u$-independent part.
\item $r_a=u^b R_{ab}$ and $r_b=u^a R_{ab}$ are the {\em left and right co-timelike parts} of $R_{ab}$. If the cotensor is symmetric, then $u^b R_{ab} = u^b R_{ba}$. 
\item $r_{a\b}= \delta_{\b}^{\;c}R_{ac}$ and $r_{\a b}= \delta_{\a}^{\;c}R_{cb}$ are the $u$-independent left and right {\em co-spacelike parts} of the $R_{ab}$ cotensor.
\end{itemize}

\section{Appendix B. Observers and Galilean transformations}
\label{App_tr}

The mathematical structure of the Galilean relativistic space-time model reflects exactly our everyday experience, that time passes independently of the observer, but the space, the environment composed by the things around us, depends on the observer. Time is absolute, space is relative. The relativity is characterized by observers. An observer is a smooth four-velocity field on the space-time (see \cite{Mat93b}), it is not necessarily global. An inertial observer is a constant four-velocity field.

Previously we have given the splitting of vectors, covectors, and tensors by a four-velocity. Absolute physical quantities are vector fields, covector fields, tensor fields, etc., that is they are vector, covector, tensor, etc. valued functions interpreted on the space-time. The splitting of the fields is local, by the local observer velocity. For example if $A^a:M\to\Mi$ is a vector field and $u^a:M\to V(1)$ is an observer, then at the world-point $x$ the $A^a(x)$ vector is reconstructed from its time- and  $u^a(x)$-spacelike parts.
\eqn{Mezfor}{
	A^a(x)=A(x)u^a(x) + A^\a(x).
}
Keeping in mind that here everything is related to a world-point, one can omit $x$ in the notation, and then our previous formulas are all valid.

Now we analyze the relation of time- and spacelike parts of the same absolute physical quantity by two  different four-velocities. In a space-time model these {\em transformation rules} can be derived.
 
In the following we need the projections by the two different four-velocities, $u$ and $u'$. Then it is simpler if we miss the explicit notation of the velocities, denoting the $u$- and $u'$-projections by $\pi^{\a}_{\;b}$ and $\pi'^{\a}_{\;\;b}$, respectively.

\subsection{Vectors}

We have seen, that the time and $u$-spacelike parts of a vector $A^a$  by an observer $u^a$ are $A=\tau_a A^a$ and $A^\a= \pi^\a_{\;b}A^b$, respectively. The non-relativistic physical theories are built on these kind of relative quantities, unaware of the deeper space-time aspects. Two different $u^a$ and $u'^a$ four-velocities may result in different time- and spacelike parts:
$$
A^a \overset{u}{\prec} \begin{pmatrix} A \cr A^\a \end{pmatrix},\qquad
A^a \overset{u'}{\prec} \begin{pmatrix} A' \cr A'^\a \end{pmatrix},
$$ 
where $\overset{u}{\prec}$ denotes the splitting by $u^a$. The $u$- and $u'$-forms of the physical quantity express the absolute four-vector with the help of its time- and spacelike parts. 

The transformation rules give the relative quantities according to an observer with the relative quantities of an other observer. In particular $A'$ and $A'^\a$ are given as a function of $A$ and $A^\a$ and the relative velocity. In our space-time model these transformation rules can be calculated by splitting the $u$-form of a physical quantity, which is $A^a=Au^a+A^\a$ in this case, by an other observer $u'^a$. Then the $u'$-timelike part of a vector 
\eqn{vi_tr}{
A' = \tau_a A^a = \tau_a (Au^a+A^\a) = A,
}
is the same, does not transform, that is Galilean invariant. This is not too surprising, because the function of the splitting, the time evaluation, does not depend on the velocities. The $u'$-spacelike part of $A^a$ is
\eqn{vt_tr0}{
A'^\a =& \pi'^\a_{\;\;b} A^b = \left(\delta^a_{\; b}-u'^a\tau_b\right) 
\left(Au^b+A^\b\right) 
= Au^a + A^\a -Au'^a = \nnl{vt_tr}
& A^\a + A(u^a-u'^a) = A^\a + Av^\a,
}
where we have denoted the relative velocity $u^a$ related to $u'^a$ as
\eqn{rels}{
v^\a =u^a-u'^a.
}
The above formula is the transformation rule of the spacelike component of a four-vector. We give it also with relative indexes
\eqn{vt_trr}{
A'^i = A^i + Av^i.
}

This is exactly the well-known {\em Galilean transformation}. \re{vt_tr} and \re{vt_trr} are the same equations written by with different notation. Let us remember, that the three-indexes $i,j,k \in \{1,2,3\}$ 
refer to the presence two four-velocities in the formula, the equation requires the presence of two observers. With this notation the absolute quantities and formulas are strictly distinguished from the usual 1+3 dimensional forms and the space-time based and Galilean transformation based ways of thinking are separated.

The complete transformation rule is 
\eqn{v_tr}{
\begin{pmatrix} A' \cr A'^i \end{pmatrix} = 
\begin{pmatrix} A \cr A^i + A v^i \end{pmatrix}.
}
In particular the transformation rule of four-velocities can be deduced directly. The splitting of a four-velocity by itself is
$$
u^a \overset{u}{\prec} \begin{pmatrix} 1 \cr 0^\a \end{pmatrix}.
$$
Therefore, the transformation rule of its spacelike part gives the expected relative velocity
\eqn{vel_id}{
\pi'^\a_{\;\;b}u^b = (\delta^a_{\; b} - u'^a\tau_b)u^b 
= u^a-u'^a = v^\a,
}
The complete transformation rule is:
\eqn{vr_tr}{
 \begin{pmatrix} 1' \cr v^i \end{pmatrix} = 
\begin{pmatrix} 1 \cr 0^i \end{pmatrix},
}
where the left hand side is $u^a$ from the point of view (in the time and space) of $u'^a$ and the right hand side is $u^a$ from the point of view (in the time and space) of itself. The meaning of this velocity transformation is that observer  $u^a$ is considered at rest according to itself, but is moving with the velocity  $v^i$ for the observer $u'^a$. Or, at the other hand, we can say  with the transformations terminology that the relativ velocity  $v^i$ for observer $u'$ is transformed to zero when we change to observer $u$. In contrast to  \re{v_tr} the dash is used only for the timelike part, because of the accustomed notation of the relative velocity ($v^i$ would be $v'^i$).

In the above formulas of Galilean transformations there are two {\em arbitrary} velocity fields, the formulas are not related solely to inertial observers. Only the Galilean invariant quantities are considered observer independent, however not only those, but a combination of properly transformed quantities may be absolute, whenever it is a component of an absolute space-time quantity. The timelike part of and absolute four-vector physical quantity (the density of an extensive quantity) is Galilean invariant, its spacelike part (the current density or flux) transforms, therefore it depends on the reference frame. However, the complete four-vector is absolute. The same is valid also for the velocity, where the four-velocity is absolute, notwithstanding that the timelike part seemingly does not contain physical information. Putting a physical theory in the space-time model one can consistently decide what depends on the reference frame and what does not.

\subsection{Covectors}

We have seen, that, the time- and spacelike parts of the covector $B_a$ by the velocity $u^a$ are $B=u^a B_a$ and $B_\a = \delta_\a^{\;b}B_b$, respectively. These componets are represented by lower indexes and horizontal mode of writing:
$$
B_a \overset{u}{\prec} (B, B_\a).
$$
According to the previous section the transformation rules of the parts of $B_a$ is obtained by splitting the $u$-form of $B_a$ using the velocity field $u'^a$. For the timelike part it is
\eqn{ki_tr}{
B' = u'^a B_a = u'^a (B\tau_a + \pi_a^{\;\b}B_\b) = B- v^\a B_\a,
}
where identity \re{vel_id} was applied. For the spacelike part one should use the  identities \re{indazon}:
\eqn{kt_tr}{
B'_\a = \delta_\a^{\;b} B_b =\delta_\a^{\;b}  (B\tau_b + \pi_b^{\;\c}B_\c) = 
B_\a.
}
The complete transformation rule may be written with in the 1+3 dimensional form:
\eqn{k_tr}{
(B', B'_i) = (B-v^i B_i, B^i).
}

A particular example is the space-time differentiation $\partial_a$, which is a symbolic covector. Then we obtain, that:
\eqn{der_tr}{
(D_{u'}, \nabla'_i) = (D_u-v^i \nabla_i, \nabla_i).
}
If $u^a$ is the velocity field of a fluid, and  $u'^a$ is that of an observer, the $D_u$ is the {\em substantial time derivative},  $v^i$ is the relative velocity of an observer, related to the fluid and  $D_{u'}= D_u-v^i \nabla_i$. The relation between the {\em partial time derivative} $D_{u'}=\partial_t$, and the substantial time derivative $D_u= d_t$ is $\partial_t= d_t -v^i\nabla_i$. It is worth to compare the previous transformation rule based derivation and the usual method (see e.g.  \cite{Gya70b}).

\subsection{Second order tensors}

An observer $u^a$ splits a second order tensor, $T^{ab}$, into a time-timelike part, $t=\tau_a\tau_b T^{ab}$, into a time-spacelike part  $t^\a=\pi^\a_{\;c}\tau_b T^{cb}$, into a space-timelike part, $t^\b=\tau_a \pi^\b_{\;c} T^{ac}$, and into a space-spacelike part, $t^{\a\b} =\pi^\a_{\;c}\pi^\b_{\;d} T^{cd}$.
Therefore
$$
T^{ab} \overset{u}{\prec} \begin{pmatrix} t & t^\a \cr t^\b & t^{\a\b} 
\end{pmatrix}.
$$
In general $t^\a \neq t^\b$ and also $t^{\a\b} \neq t^{\b\a}$, because the symmetry of $T^{ab}$ was not assumed. The corresponding transformation rules are calculated as follows. The time-timelike component is invariant
\eqn{vivi_tr}{
t' =\tau_a \tau_b T^{ab} = t.
}
The transformation rule of the time-spacelike and space-timelike components is like the transformation rule of a space vector:
\eqn{vivt_tr}{
t'^\a = \pi'^\a_{\;\;c}\tau_b T^{cb}= (\delta^a_{\;c} - u'^a\tau_c)(t 
u^c+t^\c)= t u^a-t 
u'^a+ t^\a = t^\a + t v^\a.
}
For the space-spacelike component one obtains a more complicated formula:
\eqn{vtvt_tr}{
t'^{\a\b} =& \pi'^\a_{\;\;c}\pi'^\b_{\;\;d} T^{cd}= 
\pi'^\a_{\;\;c}\pi'^\b_{\;\;d} (t u^c u^d +t^\c u^d+ 
	u^c t^\d + t^{\c\d})= \nnl{vv1}
=& t v^\a v^\b + t^\a v^\b + t^\b v^\a + t^{\a\b}.
}
Here we have used \re{vel_id}. 

The complete transformation rule can be written also with the usual notation:
\eqn{vv_tr}{
\begin{pmatrix} t' & t'^i \cr t'^j & t'^{ij} \end{pmatrix}=
\begin{pmatrix} t & t^i + tv^i \cr t^j + t v^j & t^{ij} + t^i v^j + t^j v^i + 
t v^{i}v^j\end{pmatrix}.
}

\subsection{Mixed second order tensors}

The components of a second order mixed tensor  $Q^{a}_{\;b}$ by an observer $u^a$ observer are $q=\tau_au^b Q^{a}_{\;b}$, the time-timelike,  $q^\a=\pi^\a_{\;c}u^b Q^{c}_{\;b}$, the space-timelike,  $q_\b =\tau_a\delta_\b^{\;d} Q^{a}_{\;d}$ the time-spacelike and  $q^\a_{\;\b} =\pi^\a_{\;c}\delta_\b^{\;d} Q^{c}_{\;d}$ the space-spacelike components. That is 
$$
Q^{a}_{\;b} \overset{u}{\prec} 
  \begin{pmatrix} q & q^\a \cr q_\b & q^\a_{\;\b} \end{pmatrix}.
$$
It is meaningless to speak about the symmetry of mixed second order tensors. Only , eventually, the  observer dependent split can be symmetric. This is also clear from the Galilean transformation rules, because the time-spacelike and space-timelike parts transform differently. We obtain for the time-timelike component:
\eqn{viki_tr}{
q' =\tau_au'^b Q^{a}_{\;b} =u'^b (q \tau_b + \pi_b^{\;\c}q_\c) = q- v^\c q_\c.
}
The time-spacelike component seems to be a three-vector, but does not transform. This can be understood using the identities \re{indazon}:
\eqn{vikt_tr}{
q'_\b = \tau_a\delta_\b^{\;c} Q^{a}_{\;c} =\delta_\b^{\;c} (q \tau_c + 
\pi_c^{\;\d}q_\d) = q_\b.
}
The Galilean transformation rule for the space-timelike component is:
\eqn{vtki_tr}{
q'^\a = \pi'^\a_{\;\;c} u'^b Q^{c}_{\;b} =
  \pi'^\a_{\;\;c} (q u^c + q^\c - u^cv^\d q_\d - q^\c_{\;\d}v^\d) = 
  q^\a + q v^\a - v^\a v^\b q_\b - q^\a_{\;\b}v^\b.
}
Finally for the space-spacelike part one derives 
\eqn{vtkt_tr}{
q'^{\a}_{\;\;\b} = \pi'^\a_{\;\;c}\delta_\b^{\;d}  Q^{c}_{\;d} =
\pi'^\a_{\;\;c} (q_\b u^c 
+q^\c_{\;\b}) = q^\a_{\;\b} + v^\a q_\b.
}
Here the usual identities were used. Then the complete transformation rule can be written as:
\eqn{vk_tr}{
\begin{pmatrix} q' & q'_i \cr q'^j & q'^{j}_{\;\;i} \end{pmatrix}=
\begin{pmatrix} q - v^iq_i & q_i \cr 
q^j+ v^j(q-v^kq_k)- q^j_{\;k} v^k & q^{j}_{\;i} + q_iv^j\end{pmatrix}.
}

\subsection{Second order cotensors}

The components of the second order cotensor $R_{ab}$ split by the observer $u^a$ are the $r=u^au^b R_{ab}$ time-timelike, $r_\a=\delta_\a^{\;c}u^b R_{cb}$ space-timelike, $r_\b=u^a \delta_\b^{\;d} R_{ad}$ time-spacelike and the  $r_{\a\b}= \delta_\a^{\;c}\delta_\b^{\;d} R_{cd}$ space-spacelike parts. That can be written in a matrix form as
$$
R_{ab} \overset{u}{\prec} \begin{pmatrix} r & r_\a \cr r_\b & 
r_{\a\b}\end{pmatrix}.
$$
If $R_{ab}$ is not symmetric, $r_\a \neq r_\b$ and $r_{\a\b} \neq r_{\b\a}$. The calculation of the transformation rules of the components is the following. For the time-timelike part one obtains:
\eqn{kiki_tr}{
r' = u'^au'^b R_{ab} =
u'^au'^b \left(r \tau_a\tau_b + r_\c\pi_a^{\;\c}\tau_b+ 
	r_\c\tau_a\pi_b^{\;\c} + r_{\c\d}\pi_b^{\;\c} \pi_{a}^{\;\d}\right) = 
	\nnl{kk1}
u'^a\left(r \tau_a + r_\c\pi_a^{\;\c} - r_\c\tau_a v^\c - 
	r_{\c\d}v^\c \pi_{a}^{\;\d}\right) = 
r - 2r_\c v^c + r_{\c\d}v^\c v^\d.
}
The time-spacelike part transforms identically with the space-timelike part:
\eqn{kikt_tr}{
r'_\a = \delta_\a^{\;c}u'^b R_{cb} =
  \delta_\a^{\;c} \left(r \tau_c + r_\d\pi_c^{\;\d} - r_\d\tau_c v^\d - 
  r_{\d\eh}v^\eh \pi_{c}^{\;\d}\right) = 
r_\a - r_{\a\d} v^\d.
}
The space-spacelike part is Galilean invariant
\eqn{ktkt_tr}{
r'_{\a\b} = \delta_\a^{\;c}\delta_\b^{\;d} R_{cd} = r_{\a\b}. 
}
Here the usual identities were applied, too. Finally the complete transformation rule is given in a matrix form:

\eqn{kk_tr}{
\begin{pmatrix} r' & r'_i \cr r'_j & r'_{ji}\end{pmatrix} =
\begin{pmatrix} r- 2 v^kr_k + v^kv^lr_{kl} & r_i - v^jr_{ij} \cr r_j- 
v^kt_{jk}   &  r_{ij}\end{pmatrix}.
}

\section{Acknowledgement}   

This work is based on the fundamental ideas of Tamás Matolcsi who condensed space-times into mathematical models. The author is also indebted to Tamás Fülöp, who was convinced all the time that there is an absolute nonrelativistic energy. In this work it does not look like a cotensor, but that possibility is not excluded yet. 

This work is supported by the grants Otka K104260 and K116197. 

\bibliographystyle{unsrt}

\end{document}